\newif\ifColor\Colortrue \Colorfalse
\Colortrue
\PassOptionsToPackage{super,comma,numbers,sort&compress}{natbib}  
\documentclass[%
reprint,
superscriptaddress,
 amsmath,amssymb,
 aps,
]{revtex4-2}

\usepackage{lineno}
\usepackage{graphicx}
\usepackage[utf8]{inputenc}
\usepackage{amssymb}
\usepackage{amsmath}
\usepackage{amsfonts}
\usepackage{upgreek}
\usepackage{gensymb}
\usepackage{setspace}
\usepackage{booktabs}
\usepackage[english]{babel}
\usepackage{mathtools}
\usepackage{mhchem}[version=4]
\usepackage{hyperref}
\usepackage[final]{microtype}
\usepackage{csquotes,lmodern}
\usepackage{etoolbox}
\usepackage{color}
\usepackage{multibib}
\usepackage{siunitx}
\usepackage[normalem]{ulem}
\usepackage{dcolumn}
\usepackage{bm}

\begin{document}

\title{Mapping delocalization of impurity bands across archetypal Mott-Anderson transition}
\author{M. Parzer} 
\email{michael.parzer@tuwien.ac.at}  
\author{F. Garmroudi}
\email{f.garmroudi@gmx.at}
\affiliation{Institute of Solid State Physics, Technische Universität Wien, 1040 Vienna, Austria}
\author{A. Riss}
\affiliation{Institute of Solid State Physics, Technische Universität Wien, 1040 Vienna, Austria}
\author{T. Mori}
\affiliation{International Center for Materials Nanoarchitectonics (WPI-MANA), National Institute for Materials Science (NIMS), Tsukuba, Japan}
\affiliation{Graduate School of Pure and Applied Science, University of Tsukuba, Tsukuba, Japan}
\author{A. Pustogow}
\affiliation{Institute of Solid State Physics, Technische Universität Wien, 1040 Vienna, Austria}
\author{E. Bauer}
\affiliation{Institute of Solid State Physics, Technische Universität Wien, 1040 Vienna, Austria}

\begin{abstract}
Tailoring charge transport in solids on demand is the overarching goal of condensed-matter research as it is crucial for electronic applications. Yet, often the proper tuning knob is missing and extrinsic factors such as impurities and disorder impede coherent conduction. 
Here we control the very buildup of an electronic band from impurity states within the pseudogap of ternary \ce{Fe_{2–x}V_{1+x}Al} Heusler compounds via reducing the Fe content. Our density functional theory calculations combined with specific heat and electrical resistivity experiments reveal that, initially, these states are Anderson-localized at low V concentrations $0 < x < 0.1$. As $x$ increases, we monitor the formation of mobility edges upon the archetypal Mott-Anderson transition and map the increasing bandwidth of conducting states by thermoelectric measurements. Ultimately, delocalization of charge carriers in fully disordered \ce{V3Al} results in a resistivity exactly at the Mott-Ioffe-Regel limit that is perfectly temperature-independent up to 700\,K -- more constant than constantan.

\end{abstract}
\maketitle

\onecolumngrid
\twocolumngrid

\section*{Introduction} 
The periodic arrangement of atoms in crystals forms energy bands, where electrons behave as extended Bloch waves. In single atoms, however, electrons occupy localized energy levels. It was Sir Nevill Mott, who first approached the fundamental question in condensed matter physics  of band formation, transitioning from localized insulating states to delocalized metallic states. He argued that as two atoms move closer, at a certain point, a critical threshold is reached where the screening of ions by neighboring electrons becomes strong enough to delocalize them \citep{mott1956transition}.

\begin{figure}[h!]
\includegraphics[width=0.48\textwidth]{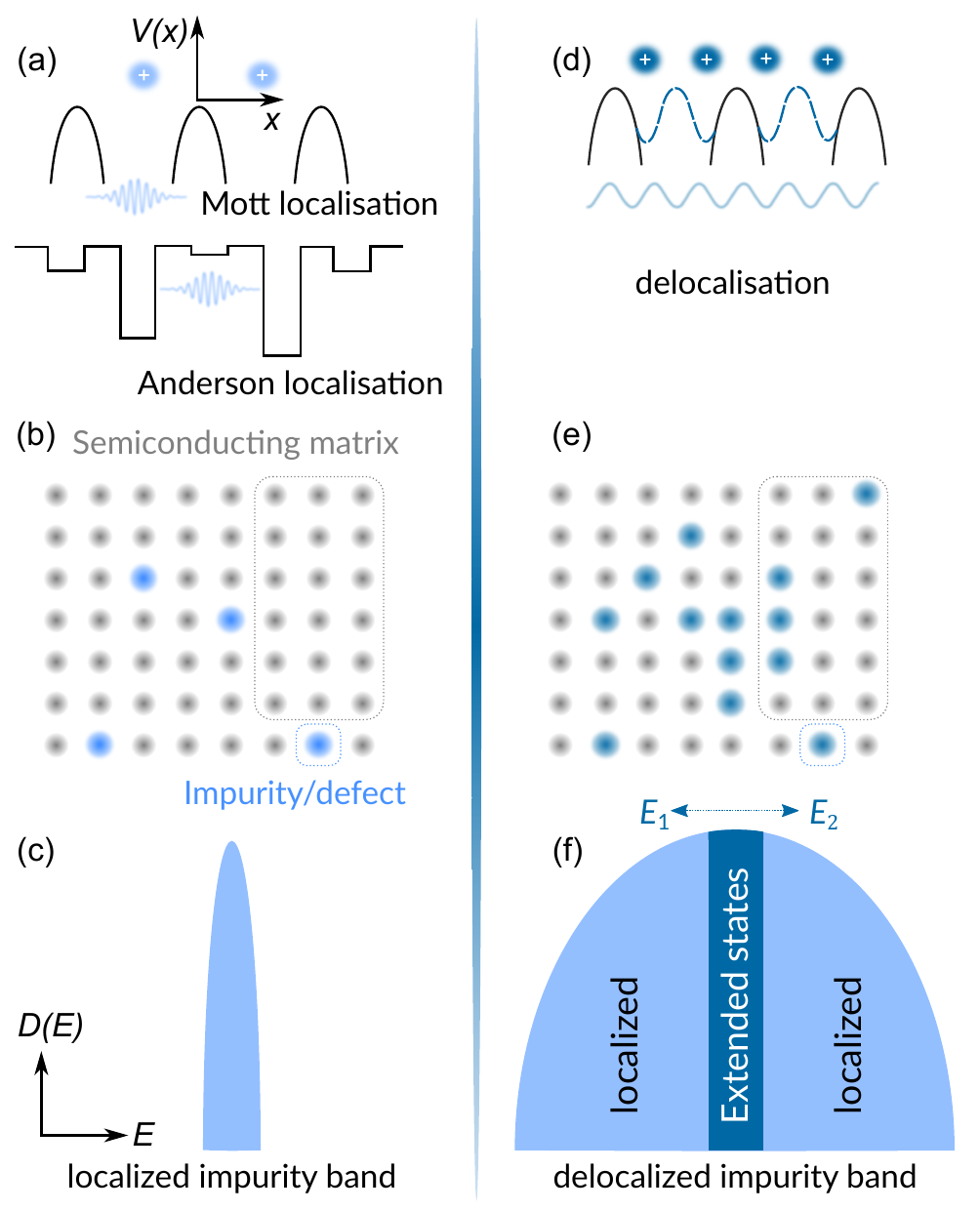}
\caption{\textbf{Sketch of the Mott-Anderson transition.} (a) Mott-type and Anderson-type charge localization mechanisms in solids. (b) Semiconducting matrix with isolated impurities and (c) a corresponding localized IB. (d) Atomic potentials, hosting extended Bloch states. (e) Semiconducting matrix with percolation of impurities and (f) a partly delocalized IB.}
\label{fig:Fig1}
\end{figure}

In doped semiconductors, a similar, but more realistic, transition occurs. At low doping levels, electrons remain trapped at impurity sites, but increasing dopant concentration leads to a critical concentration, $x_\text{c}$, where an insulator-metal phase transition occurs \citep{mott1956transition,mott1961theory}. Additionally, correlations among the electrons can localize and split the narrow impurity bands (IBs) by an energy gap $\Delta = U-W$, with $U$ being the Coulomb interaction and $W$ the band width.
In parallel to Mott’s work, Anderson developed a theory of disorder-induced localization of electrons in matter due to their wave-like nature \cite{Anderson1958}. In its essence, Anderson localization describes the absence of diffusion of wave-like objects due to the multiple scattering events and quantum interference of self-intersecting scattering paths \cite{cutler1969observation,schwartz2007transport,
roati2008anderson,dobrosavljevic2012conductor,segev2013anderson}. In low-doped semiconductors, both Mott and Anderson localization -- the latter arising from the inherently random distribution of impurity atoms on the ordered crystal lattice -- are believed to contribute to the insulating nature of the IB \cite{mott1956transition,belitz1994anderson}.  Fig.\ref{fig:Fig1} illustrates both these localization mechanisms and the way in which delocalization has been theoretically predicted to take place: At $x_c$, two mobility edges $E_\text{c}$ emerge in the center of the IB. These critical energies separate localized states in the band tails from delocalized ones in the center \cite{mott1967electrons}.

The delocalization of IBs and their effect on electronic transport remain an important question, given the widespread use of dopant atoms with in-gap impurity states to control the physical properties of semiconductors in various electronic devices \cite{prati2012anderson,yu2014impurity, garmroudi2022anderson, ho2024quantum}. However, up until now, extensive investigation of the Mott-Anderson transition (MAT) in IBs has frequently been hindered by the low solubility of dopant atoms and the fact that the IBs usually hybridize with the bulk conduction bands. This has hindered the unambiguous analysis of the MAT in IBs, resulting, for instance, in the so-called exponent puzzle for the critical exponent describing the universality of the transition \citep{stupp1993possible,carnio2019resolution}. 

\begin{figure*}[t!]
\includegraphics[width=1\textwidth]{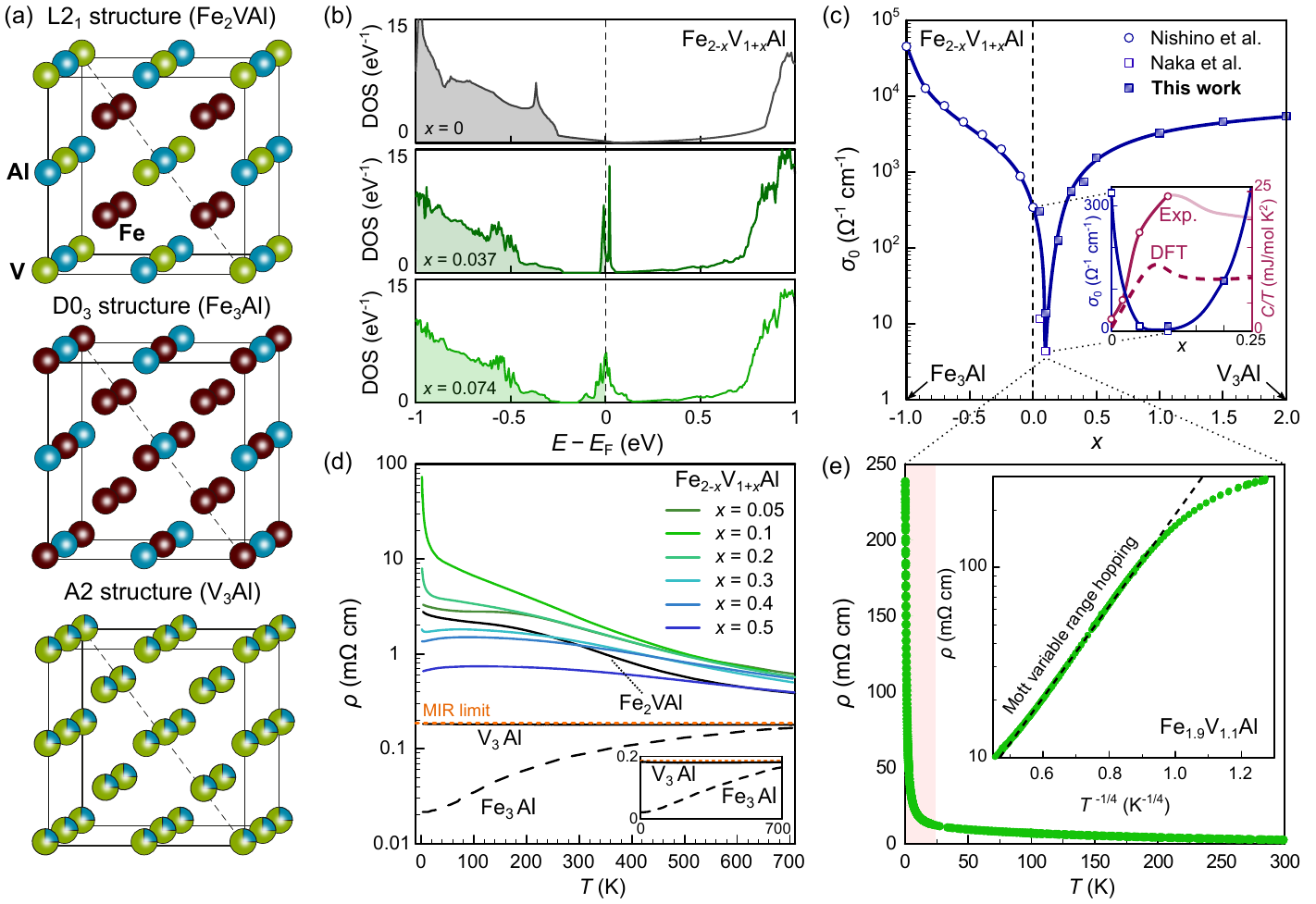}
\caption{\textbf{Structure and electronic transport of \ce{Fe_{2-x}V_{1+x}Al} system.} (a) Crystal structures of ternary Heusler compound \ce{Fe2VAl} crystallizing in the fully ordered L2$_1$ structure, binary \ce{Fe3Al} in the partly disordered D0$_3$ structures and \ce{V3Al} in the fully-disordered A2 structure. (b) Densities of states of \ce{Fe_{2-x}V_{1+x}Al} from DFT supercell calculations with $x=0$, 0.037 and 0.074. (c) Composition-dependent electrical conductivity at $T=2\,$K of \ce{Fe_{2-x}V_{1+x}Al} over the whole compositional phase space, from $-1 \leq x \leq 2$. A distinct minimum is observed at $x=0.1$. Inset shows a magnified view of the conductivity minimum alongside the density of states at the Fermi level derived from specific heat data \cite{naka2016composition} and our DFT calculations, both showing a maximum instead. (d) Temperature-dependent resistivity $\rho(T)$ of \ce{Fe_{2-x}V_{1+x}Al} close to the significant minimum ($x=0.1$--$0.5$), together with the two endpoints of the composition transition. Notably, \ce{Fe3Al} shows metallic behavior, while the resistivity of \ce{V3Al} is completely flat, with a residual resistivity ratio of $\text{RRR}<1.015$ which is right at the Mott-Ioffe-Regel (MIR) limit (dotted red line), to which \ce{Fe3Al} converges at high temperatures. The inset shows $\rho(T)$ of the binary compounds on linear scale. (e) Signatures of Mott-Anderson charge localization in $\rho(T)$ of \ce{Fe_{1.9}V_{1.1}Al}, measured down to 0.39\,K. An extremely sharp upturn of $\rho(T)$  takes place below $T\approx 20\,$K, consistent in with Mott variable range hopping (inset) in a broad range of temperatures.}
\label{fig:Fig2}
\end{figure*}

Here, we successfully map the delocaliatzion of IBs in semiconducting \ce{Fe_{2-x}V_{1+x}Al} Heusler compounds. This system offers exceptional tunability of the chemical composition ($-1 \leq x\leq 2$) within a similar cubic structure, ranging from \ce{Fe3Al} with the partially disordered D0$_3$ structure up to \ce{V3Al} with the fully disordered A2 structure (see Fig.\,\ref{fig:Fig2}\,a). Crucially, the IB arises exactly within the center of the gap and its band width remains remarkably narrow up to large concentrations of $x$. This confines the electronic transport at low temperatures to the IB states, allowing to map the MAT from the fully insulating to the disordered and correlated metal regime.

The Heusler compound \ce{Fe2VAl}, crystallizing in the fully ordered L2$_1$ structure, has long been known as a promising thermoelectric material for near-ambient temperature applications \cite{pecunia2023roadmap,nishino2006thermal,mikami2008development,mikami2012thermoelectric}, owing to its narrow pseudogap in the electronic structure. Over the recent years, various substitution studies have been performed to optimize the position of the Fermi energy and to tailor the band structure itself, resulting in outstanding thermoelectric power factors that rival or exceed those of state-of-the-art \ce{Bi2Te3} thermoelectrics \cite{miyazaki2013thermoelectric,garmroudi2021boosting,garmroudi2022large,
alleno2023optimization}.
 
A highly unconventional doping dependence of the thermoelectric properties has been reported for off-stoichiometric, self-substituted \ce{Fe_{2-x}V_{1+x}Al} \cite{hanada2001seebeck,nishino2014doping}. Fig.\,S1 in the Supplemental Materials compares the Seebeck coefficient $S$ as a function of the valence electron concentration per atom (VEC) for \ce{Fe_{2-x}V_{1+x}Al} and various substituted \ce{Fe2VAl}-based compounds with different substitution elements. Typically, for $\text{VEC} < 6$ (corresponding to $p$-type doping) $S$ is expected to be positive, and for $\text{VEC} > 6$ ($n$-type doping) $S<0$ is expected. This trend is indeed observed in all known substituted compounds except for \ce{Fe_{2-x}V_{1+x}Al}, where an entirely opposite behavior emerges.   
Moreover, Naka et al. reported a composition-induced metal--insulator quantum phase transition in V-rich \ce{Fe_{2-x}V_{1+x}Al} and a ferromagnetic quantum critical point in Fe-rich \ce{Fe_{2-x}V_{1+x}Al} \cite{naka2012ferromagnetic,naka2016composition}, yet the exact mechanisms behind these transitions remain unclear.
Here we demonstrate that the origin of these quantum phase transitions and the highly unconventional doping behavior can be traced back to the delocalization of narrow IBs right next to $E_\text{F}$. Crucially, using temperature-dependent thermoelectric transport measurements we quantitatively map the width of the delocalized IB as a function of the impurity concentration.

\begin{figure*}[t!]
\centering
\includegraphics[width=1\textwidth]{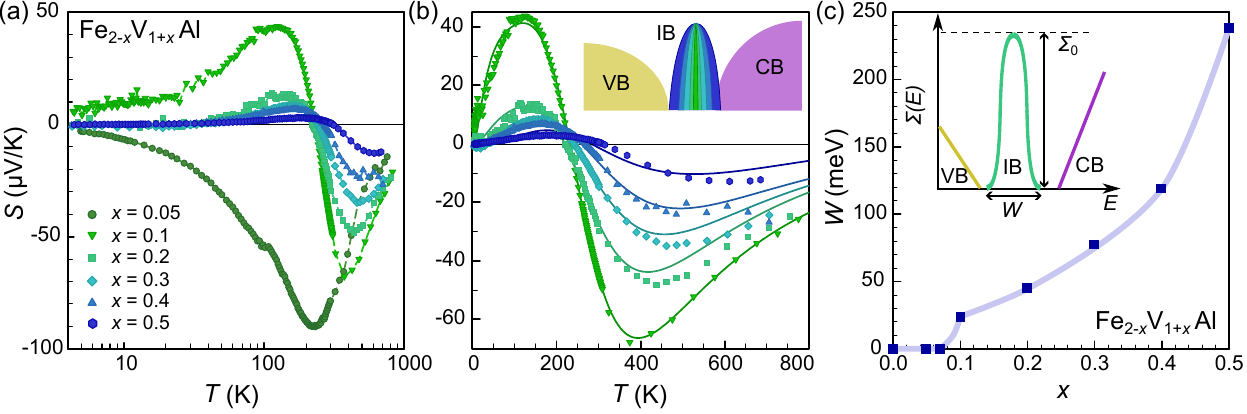}
\caption{\textbf{Mapping delocalization of impurity bands via thermoelectric transport.}(a) Temperature-dependent Seebeck coefficient $S(T)$ of \ce{Fe_{2-x}V_{1+x}Al} with $x=0.05-0.5$. For $x=0.05$, the impurities merely act as fully localized donor states, placing $E_\text{F}$ right below the dispersive conduction band (see Fig.\,\ref{fig:Fig2}\,b) and leading to a negative $S(T)$ over the entire temperature range. As $x$ increases up to 0.1, a distinct positive maximum develops around 120\,K, which gets progressively smeared out and shifted towards higher temperatures with increasing $x$. (b) Experimental $S(T)$ and least-squares fits (solid lines) employing a three-band model with parabolic valence and conduction bands and a delocalized impurity band confined to a width $W$. (c) Quantitative mapping of the composition dependence of $W$ with a progressive broadening at $x>0.1$. Inset shows a sketch of the energy-dependent transport distribution function $\Sigma(E)$ of our model.}
\label{fig:Fig3}
\end{figure*}
To obtain insight into the states around $E_\text{F}$ relevant for charge transport, we performed density functional theory (DFT) calculations on supercells of substituted \ce{Fe_{2-x}V_{1+x}Al} with varying concentration of defects. Fig.\,\ref{fig:Fig2}\,b displays the density of states (DOS) around $E_\text{F}$ for $x=0,\,0.037$ and 0.074. It is evident that for $x=0\rightarrow 0.037$, narrow and highly localized impurity states emerge within the pseudogap at the lower edge of the dispersive conduction band (CB). As the concentration of V antisites on the two Fe sublattices continues to increase, these states gradually broaden, eventually filling the entire pseudogap and transforming the system into a more metallic-like state. However, the periodic boundary conditions (as per Bloch's theorem) imposed by the supercell approach and the lack of accounting for the decoherence of the wave functions prevent the accurate description of the Anderson-localized nature of these states. Thus, simple DFT calculations can only predict that there exist such impurity states but not whether and to which extent they are localized or not.

On the other hand, examining composition- and temperature-dependent electronic transport of \ce{Fe_{2-x}V_{1+x}Al}, displayed in \autoref{fig:Fig2}\,c-e, reveals clear evidence for the Anderson-localized nature of these IBs as the low-temperature conductivity drops by several orders of magnitude and exhibits a sharp minimum at $x\approx 0.1$ (see Fig.\,\ref{fig:Fig2}\,c). The inset in Fig.\,\ref{fig:Fig2}\,c compares the composition-dependent DOS at the Fermi level $D(E_\text{F})$ obtained from DFT and specific heat data ($C/T$)\,\cite{naka2016composition} with the conductivity at $T=2\,$K. While, both DFT calculations and $C/T$ data reveal a peak at $x=0.1$ associated with the IBs, $\sigma_{2K}$ vanishes and only starts to increase at $x>0.1$ when the impurities start to form a "real", delocalized band. 

Temperature-dependent resistivity curves for several different compositions across a wide range of temperatures and compositional phase space are presented in Fig.\,\ref{fig:Fig2}\,d. Focusing first on the end compounds, the magnetic metal \ce{Fe3Al} exhibits a metallic-like resistivity, $d\rho/dT>0$, well described by the Bloch-Wilson law for $T<T_\text{C}$ (see SM), with $T_\text{C}\approx 740\,$K. On the other hand, \ce{V3Al} with the fully disordered A2 structure (see Ref.\,\cite{zhang2021pauli}), shows a negligible, negative temperature dependence of $\rho(T)$ with a residual resistivity ratio $\text{RRR} < 1.015$, aligning with the phenomenological Mooij criterion \citep{mooij1973electrical, disante2017disorder, ciuchi2018origin} (see also Fig. S2 of the Supplementary Materials). Note that $\rho(T)$ of \ce{Fe3Al} saturates towards the value of \ce{V3Al}, which lies almost exactly at the Mott-Ioffe-Regel (MIR) limit (red dotted line), even in a non-logarithmic plot (inset Fig.\,\ref{fig:Fig2}\,d).. The MIR limit states that the charge carrier mean free path is limited by the interatomic distance $k_\text{F}\,l_\text{min} \sim 2\pi$, where $k_\text{F}$ is the Fermi wave vector and $l_\text{min}$ the minimal mean free path. For spherical Fermi surfaces, Calandra and Gunnarsson derived an expression \cite{calandra2002electrical,gunnarsson2003colloquium}

\begin{equation}
\rho_\text{MIR}=\frac{3\pi^2\hbar}{e^2\,k_\text{F}^2\,l_\text{min}}\;.
\end{equation}

\noindent Assuming $k_\text{F} \approx  \pi/a$ and using the experimental lattice parameter of \ce{V3Al}, this yields $\rho_\text{MIR} \approx 187\,\upmu \Omega$cm (red dotted line in Fig.\,\ref{fig:Fig2}\,d), which almost perfectly coincides with the temperature-independent resistivity value of \ce{V3Al} $\rho(T) \approx \text{const.} \approx 180\,\upmu \Omega$cm up to 700\,K. Notably, $\rho(T)$ of this fully disordered metal is more constant than in constantan.

In the ternary system, a deep pseudogap develops around the Fermi energy for \ce{Fe2VAl}, which leads to a semiconductor-like resistivity, $d\rho(T)/dT$ that can be mainly attributed to the thermal activation of charge carriers across the narrow pseudogap \cite{okamura2000pseudogap,nishino2001electronic}, although recently it has been shown that charge carrier scattering may also play an important role owing to residual intrinsic antisite exchange defects \cite{garmroudi2023pivotal}. Notably, $\rho(T)$ increases further by more than an order of magnitude when additional V atoms are substituted on the Fe sites in \ce{Fe_{1.9}V_{1.1}Al}, with a  pronounced upturn at the lowest temperatures. Fig.\,\ref{fig:Fig2}\,e shows a linear plot of $\rho(T)$ of \ce{Fe_{1.9}V_{1.1}Al} with measurements down to $T=0.39\,$K. The extremely sharp increase of $\rho(T)$ below $\approx 20\,$K is consistent with the presence of Anderson-localized states around $E_\text{F}$ (see inset Fig.\,\ref{fig:Fig2}\,e), although at the lowest temperatures, $T\lesssim 1.2$\,K, $\rho(T)$ deviates significantly from variable range hopping behavior and can instead be described phenomenologically by a simple power law $\rho(T)\propto T^{-\alpha}$, with $\alpha=1.3-1.4$, consistent with what has been reported by Naka et al. \cite{naka2016composition} and also, e.g., in doped silicon near the MAT \cite{mott1983metal, long1985magnetic, shafarman1989dc}.

At low temperatures, when thermally activated hopping of electrons from one localized impurity to the next, which are far apart in real space but close in energy, dominates, a temperature dependence $\rho(T) \propto \text{exp}\,\left[\left(T_0/T\right)^{1/4}\right]$ is expected in three dimensions -- commonly known as Mott variable range hopping (VRH). Here, $T_0$ is the characteristic temperature, which depends inversely on the localization length 

\begin{equation}
\xi_\text{L}=\left(\frac{1}{18}D(E_\text{F})\,k_\text{B}\,T_0\right)^{-1/3}\;.
\end{equation}

\noindent The localization length indicates the exponential decay of the wave function $|\Psi(r)|^2 \sim \text{exp} \left(-|r-r_0|/\xi_\text{L}\right)$ at an impurity site $r_0$. By fitting the slope of $\ln\,\rho$ versus $T^{-1/4}$ of \ce{Fe_{1.9}V_{1.1}Al} at low temperatures, $T_0$ is obtained as $\approx 490$\,K. Furthermore, using $D(E_\text{F})$, derived from $C/T$ data from Ref.\,\cite{naka2016composition}, this leads to a localization length $\xi_\text{L}\approx 12$\,nm, equaling about 21 unit cells. If one considers that for \ce{Fe_{1.9}V_{1.1}Al} about one impurity is present in every three unit cells, this suggests that already a significant overlap of the Anderson-localized wave functions exists. As will be demonstrated below, using temperature-dependent thermoelectric measurements, we confirm the existence of a delocalized regime of the IB and quantitatively assess its bandwidth $W$, which is not possible by merely investigating $\rho(T)$.

Fig.\,\ref{fig:Fig3} shows $S(T)$ of \ce{Fe_{2-x}V_{1+x}Al} with $0.05 \leq x\leq 0.5$. For $x=0.05$, the impurity states remain fully localized and do not actively contribute themselves to the transport properties. Instead, $S(T)$ is governed by the dispersive CB states up to 230\,K, where holes from the dispersive valence band (VB) states are thermally activated and bipolar conduction takes over (see also SM). As $x\geq 0.1$, a pronounced positive peak of $S(T)$ develops at low temperatures, which we assign to delocalization of the IB. To model the contribution of a (partly) delocalized IB to the Seebeck coefficient, a band with a finite width $W$ was integrated into the two-parabolic band modeling framework commonly employed to understand charge transport in pristine \ce{Fe2VAl} and conventional doping scenarios \cite{anand2020thermoelectric, garmroudi2021boosting, hinterleitner2021electronic}. For parabolic bands, the energy-dependent transport function $\Sigma(E)= D(E)v(E)^2\tau(E)$ increases linearly at the band edge $\Sigma(E)\propto E$ \cite{zevalkink2018practical}. 

For the delocalized IB with finite width $W$, we developed a model in which $\Sigma(E)$ increases at either side of the mobility edge with

\begin{align}
\Sigma(E,T) &= \Sigma_0(T)\,\left(\frac{E-E_c}{k_\text{B}T}\right)^\nu\;.
\label{eq:trans_func_IB}
\end{align}

\noindent Here, $\nu$ is the critical exponent of the Anderson transition that can vary between 0.5 and 2, depending on compensation and band hybridization \cite{carnio2019resolution}. In our model, the entirety of the IB can be engineered using expressions following \autoref{eq:trans_func_IB} to yield a single continuously differentiable function (for details see SM). Finally, the contribution of the entire IB to the transport properties can be evaluated from three independent parameters: (i) its bandwidth $W$, its maximum $\Sigma_0$ and (iii) the position of $E_\text{F}$.

$\Sigma(E)$ of the parabolic bands and the delocalized IB are then used to numerically solve the transport integrals and calculate the Seebeck coefficient

\begin{equation}
S(T) = \frac{k_\text{B}}{e\,T}\frac{\int_{-\infty}^{\infty}\Sigma(E)\left(E-\mu\right)\left(-\partial f/\partial E\right)\,dE}{\int_{-\infty}^{\infty}\Sigma(E)\left(-\partial f/\partial E\right)\,dE}\;,
\label{eq:Seebeck}
\end{equation}

\noindent with $f(E,\mu,T)$ and $\mu(T)$ being the Fermi-Dirac distribution and the chemical potential, respectively; the numerator in \autoref{eq:Seebeck} represents the electrical conductivity. Since multiple bands contribute to the transport properties, the single-band contributions $S_i$ have to be weighted with their respective conductivities $\sigma_i$ 

\begin{equation}
S_\text{tot}=\sum_i \frac{S_i\,\sigma_i}{\sigma_i}\;,
\label{eq:Stot}
\end{equation}

\noindent where $i=\lbrace \text{VB, CB, IB}\rbrace$. \autoref{eq:Stot} was fitted to the experimental $S(T)$ data in the following way: Firstly, \ce{Fe_{1.9}V_{1.1}Al}, which exhibits the most significant temperature dependence in $S(T)$, was fitted, yielding excellent agreement (see Fig.\,\ref{fig:Fig3}\,b). Here, the fit parameters are the two energy gaps between the IB center and the VB and CB, $\Delta_\text{VB}$ and $\Delta_\text{CB}$ the effective mass ratio between the CB and VB, $m_\text{CB}/\text{VB}$, as well as $W$, $\Sigma_0$ and the position of $E_\text{F}$ with respect to the impurity band center. In a second step, all these model parameters except for the delocalized IB width $W$ were fixed. Then, all the measured curves for the remaining samples with $x>0.1$ were modelled in a single-parameter fit and  perfectly reproduced, highlighting the robustness of our modelling approach and the obtained parameter values.

With respect to the IB center, the energy gaps were determined as $120\,$meV towards the valence band and $100\,$meV towards the conduction band edge, highlighting that the impurity band forms in the middle of the gap, in agreement with our DFT results. The effective mass ratio of the two bands $m_\text{CB}/m_\text{VB}$ is obtained as $\approx  2$, in good agreement with previous parabolic band modelling analyses of \ce{Fe2VAl}-based thermoelectrics \cite{anand2020thermoelectric}. Most importantly, for \ce{Fe_{1.9}V_{1.1}Al}, our analysis reveals an extremely narrow region ($W \approx 24$\,meV) of the IB and $E_\text{F}$ positioned right next to the center of the IB, in striking agreement with the DFT-derived DOS. As $x$ increases, our modelling scheme accurately predicts the delocalization of the IB as $W$ progressively increases up to 240\,meV in \ce{Fe_{1.5}V_{1.5}Al} as shown in \autoref{fig:Fig3}\,c.

In conclusion, our study provides a thorough investigation of the \ce{Fe_{2-x}V_{1+x}Al} Heusler system, focusing on the composition-induced Mott-Anderson transition. By examining a wide range of compositions from \ce{Fe3Al} to \ce{V3Al}, we identified a profound conductivity minimum in the V self-substituted Heusler compound \ce{Fe_{1.9}V_{1.1}Al}. Temperature-dependent resistivity measurements near this stoichiometry reveal Mott variable range hopping behavior with a localization length $\xi_\text{L}\approx 12$\,nm. The temperature-dependent Seebeck coefficient was measured over a wide range of temperatures and compositions and a charge transport model was developed to accurately describe thermoelectric transport in delocalized impurity bands. Analyzing various $S(T)$ data sets within this modelling framework, we successfully capture and quantitatively map the delocalization of IBs in \ce{Fe_{2-x}V_{1+x}Al}. Our findings emphasize the value of thermoelectric transport measurements in probing the electronic structure and transport mechanisms in complex materials, especially when paired with appropriate multi-band fitting models. This work not only advances the understanding of \ce{Fe2VAl}-based thermoelectrics and solves the issue of the unconventional doping behavior in \ce{Fe_{2-x}V_{1+x}Al}, but also provides a robust framework for studying similar transitions in other materials.

\textit{Acknowledgements.--} The authors thank V. Dobrosavljevic and S. Fratini for fruitful discussions. Financial support for M.P., F.G., A.R., T.M. and E.B. came from the Japan Science and Technology Agency (JST), program MIRAI, JPMJMI19A1. The computational results presented have been achieved using the Vienna Scientific Cluster (VSC).

\end{document}